                      \def\version{3 August 2007}               %

\documentclass[reqno,twoside,11pt]{amsart}

\usepackage{amsmath}
\usepackage{amssymb}
\usepackage{bm}

\usepackage{graphicx}

\def\d{\delta}

\newfam\Bbbfam
\font\tenBbb=msbm10 \font\sevenBbb=msbm7 \font\fiveBbb=msbm5
\textfont\Bbbfam=\tenBbb \scriptfont\Bbbfam=\sevenBbb
\scriptscriptfont\Bbbfam=\fiveBbb

\newcommand{\C}     {\mathbb{C}}
\newcommand{\R}     {\mathbb{R}}
\newcommand{\Z}     {\mathbb{Z}}
\newcommand{\N}     {\mathbb{N}}

\def\1{{\mathchoice {1\mskip-4mu\mathrm l}      
{1\mskip-4mu\mathrm l} {1\mskip-4.5mu\mathrm l}
{1\mskip-5mu\mathrm l}}}

\def\comment#1{}
\newtheoremstyle{thm}{2ex}{2ex}{\itshape\rmfamily}{}
{\bfseries\rmfamily}{}{1.7ex}{}

\newtheoremstyle{rem}{1.3ex}{1.3ex}{\rmfamily}{}
{\itshape\rmfamily}{}{1.5ex}{}

\newenvironment{proofsect}[1]
{\vskip0.1cm\noindent{\bf #1.}\hskip0.5cm}

\newcommand\Tr{\mathop{\rm Tr}\nolimits}

\newcommand\half{\frac{1}{2}}

\newcommand\supp{\mathop{\rm supp}\nolimits}

\newcommand\eps{\epsilon}

\newcommand\be{\begin{equation}}
\newcommand\ee{\end{equation}}
\newcommand\bea{\begin{eqnarray}}
\newcommand\eea{\end{eqnarray}}
\newcommand\non{\nonumber}

\newtheorem{theorem}{Theorem}[section]
\newtheorem{lemma}[theorem]{Lemma}

\theoremstyle{definition}

\newtheorem{bem}[theorem] {Remark}

\renewcommand{\section}{\secdef\sct\sect}
\newcommand{\sct}[2][default]{\refstepcounter{section}
\vspace{0.8cm} \setcounter{equation}{0}
\centerline{ 
\large\scshape \arabic{section}.\ #1} \vspace{0.2cm}}
\newcommand{\sect}[1]{
\vspace{0.8cm} \centerline{\large\scshape #1} \vspace{0.2cm}}

\renewcommand{\subsection}{\secdef \subsct\sbsect}
\newcommand{\subsct}[2][default]{\refstepcounter{subsection}
\nopagebreak \vspace{0.5\baselineskip} {\flushleft\bf
\arabic{section}.\arabic{subsection}~\bf #1  } \nopagebreak}
\newcommand{\sbsect}[1]{\vspace{0.1cm}\noindent
{\bf #1}\vspace{0.1cm}}
\newcommand{\heap}[2]{\genfrac{}{}{0pt}{}{#1}{#2}}

\renewcommand{\subsubsection}{%
\secdef \subsubsect\sbsbsect}
\newcommand{\subsubsect}[2][default]{%
\refstepcounter{subsubsection} \nopagebreak
\vspace{0.1\baselineskip} \nopagebreak {\flushleft
\sffamily\slshape
\arabic{section}.\arabic{subsection}.\arabic{subsubsection}
\ %
\sffamily #1\/.}\ }
\newcommand{\sbsbsect}[1]{\vspace{0.1cm}\noindent
{\bf #1}\ }

\renewcommand{\d}{{\rm d}}

\newcommand{\Acal}   {{\mathcal A }}

\newcommand{\Scal}   {{\mathcal S }}

\begin{document}

\title[Bose-Hubbard model]{\large C$^*$-algebraic approach to the Bose-Hubbard model}

\author[Stefan Adams and Tony Dorlas]{} 

\date{}

\maketitle

\thispagestyle{empty}
\vspace{0.2cm}
\centerline {\sc By Stefan Adams\footnote{Max-Planck Institute for Mathematics in the Sciences, Inselstra{\ss}e 22-26, D-04103 Leipzig, Germany, Dublin Institute for Advanced Studies, School of Theoretical Physics, 10, Burlington Road, Dublin 4, Ireland, {\tt adams@mis.mpg.de}} \/ and  Tony Dorlas\footnote{Dublin Institute for Advanced Studies, School of Theoretical Physics, 10, Burlington Road, Dublin 4, Ireland, {\tt dorlas@stp.dias.ie}}}
\vspace{0.4cm}
\renewcommand{\thefootnote}{}
 \footnote{
Partially supported by DFG Forschergruppe 718 \textit{Analysis and Stochastics in Complex Physical Systems
 Berlin and Leipzig}.}
\maketitle

\thispagestyle{empty} \vspace{0.2cm}

\vspace{0.2cm}

\centerline{\small(\version)} \vspace{.3cm}
\begin{quote}
{\small {\bf Abstract:}} We give a new derivation of the variational formula for the
pressure of the long-range-hopping Bose-Hubbard model, which was
first proved in \cite{BD}. The proof is analogous to that of a
theorem on noncommutative large deviations introduced by Petz,
Raggio and Verbeure \cite{PRV} and could similarly be extended to
more general Bose system of mean-field type. We apply this
formalism to prove Bose-Einstein condensation for the case of
small coupling.
\end{quote}
\vfill


\vfill
\eject

\setcounter{section}{0}
\section{Introduction}

The Bose-Hubbard Model with nearest-neighbour hopping is defined
on a cubic lattice $\Z^d$ of dimension $d$ by the Hamiltonian
 \be   H = \half \sum_{x,y \in \Z^d:\, |x-y|=1} (a_x^* -
a_y^*)(a_x - a_y) + \lambda \sum_x n_x (n_x-1), \ee where the
operators $a_x^*$ and $a_x$ are creation and annihilation
operators satisfying the usual bosonic commutation relations
$$ [a_x, a_y^*] = \delta_{x,y}. $$ The operators $n_x = a_x^* a_x$
are the local number operators.

In \cite{BD} a long-range-hopping version of this model was
analysed. It is given by the Hamiltonian \be  H_V = \frac{1}{2V}
\sum_{x,y=1}^V (a_x^* - a_y^*)(a_x - a_y) + \lambda \sum_x n_x
(n_x-1) \ee on a complete graph of $V$ sites. In particular, the
following variational expression for the pressure was derived:

\begin{equation}\label{pformula} 
\begin{aligned}
p(\beta,&\mu,\lambda)=\\
& \sup_{r \geq 0} \Big\{- r^2 + \frac{1}{\beta} \ln \Tr \exp \Big[
\beta \Big((\mu+\lambda-1)n - \lambda n^2 + r(a + a^*) \Big)
\Big] \Big\}.
\end{aligned}
\end{equation} The derivation made
use of the so-called approximating Hamiltonian method introduced
by Bogoliubov Jr. \cite{Bog} and made rigorous by Zagrebnov et al.
\cite{Zetal}, see also \cite{BZ}. Here we present a new derivation
of this formula using the C$^*$-algebraic method of Petz, Raggio
and Verbeure \cite{PRV}, which is inspired in part by Varadhan's
theorem in probabilistic large deviation theory, and in part by
work of Fannes, Spohn and Verbeure \cite{FSV}. Our C$^*$-algebraic 
approach corresponds to variational expressions for thermodynamic 
functionals of classical Gibbs measures \cite{georgii}, and in particular we are using the variational expression for the relative entropy (see \cite{DS01} for classical probability theory). As a result we obtain a variational expression for the pressure where we optimize over states of the infinite system. Due to the symmetry (lack of geometry) on the complete graph this expression can be simplified to \eqref{pformula} using the well-known St{\o}rmer theorem.  Hence, our results here are a quantum analogy of recent results for classical statistical mechanical models on complete graphs using exchangeability \cite{Liggett}.
The analysis in
\cite{PRV} concerns quantum spin models and the extension to the
Bose-Hubbard model requires a number of technical considerations.
Many of these can be found in the book by Ohya and Petz \cite{OP},
but we present some of the proofs here nonetheless in order to
make this paper more self-contained.

Section~\ref{sec-2} contains the proof of our variational formula \eqref{pformula}. In Section~\ref{sec-3} we briefly present
some of the features of the model again, mainly in order to
correct some minor but irritating errors in the analysis of
\cite{BD}. In Section~\ref{sec-4}, we show that some of the C$^*$-algebraic
formalism extends to the nearest-neighbour hopping model,
resulting in a variational formula for the pressure analogous to
the well-known formula for spin models. This formula does not
appear to have been written down before. As the variation is over
the set of all translation-invariant states on the lattice, it is
difficult to analyse, however, just as in the case of spin models.

\section{C$^*$-algebraic derivation}\label{sec-2}

We follow the technique of Petz, Raggio and Verbeure \cite{PRV}. 
The Hamiltonian is invariant under the permutation group. We write
it as 

\begin{equation}\label{Hamiltonian} H_V = -V \Big(\frac{1}{V} \sum_{x=1}^V a_x^* \Big)
\Big(\frac{1}{V} \sum_{y=1}^V a_y \Big) + \sum_{x=1}^V h_x,
\end{equation}
where 
$$
h_x = n_x + \lambda n_x(n_x-1). 
$$ For finite $V$
we thus assume that a CCR algebra $ \Acal_V $ is given generated
by creation and annihilation operators $a_x^*$ and $a_x$ ($x \in
\{1,\dots,V\}$) and with standard representation on the Fock space
${\mathcal  F}_V$.

Next we define a  reference state $\omega_V$ on the Fock space
${\mathcal F}_V$ as the product state $\omega_V = \bigotimes_{x=1}^V
\omega_x$, where $\omega_x$ has the density matrix 
$$
\rho_{\omega_x} = \frac{\exp [\beta(\mu n_x - h_x)]}{\Tr \exp
[\beta(\mu n_x - h_x)]}. 
$$

Let 
\begin{equation} 
{\mathcal  A} = \overline{\bigcup_{V=1}^\infty {\mathcal  A}_V} 
\end{equation}
be the (quasi-local) CCR algebra for the complete lattice $\N$, where the closure is taken with respect to the norm topology.

\begin{lemma}\label{meanenergy} Suppose that $\phi$ is a regular permutation-invariant
state on $\mathcal  A$ such that $\phi(n_x) < +\infty$ for all $x\in\{1,\ldots,V\}$.
Then $$ \lim_{V \to \infty} \phi \Big( \Big(\frac{1}{V}
\sum_{x=1}^V a_x^* \Big) \Big( \frac{1}{V} \sum_{y=1}^V a_y
\Big) \Big) = \phi(a_1^* a_2). $$ \label{L1} \end{lemma}

\begin{proofsect}{Proof} Notice that we can define $\phi(n_x)$ as the
supremum $$\phi(n_x) = \sup_{N\ge 1}\,\phi(P_N^{(x)} n_x), $$ where
$P_N^{(x)}$ is the projection on the subspace of ${\mathcal  F}_x$ with
$n_x \leq N$. Obviously, this is independent of $x$ by permutation
invariance. The formula $\phi \big(\big(\frac{1}{V}
\sum_{x=1}^V a_x^* \big) \big( \frac{1}{V} \sum_{y=1}^V a_y
\big)\big)$ should be interpreted in a similar way:
$$ 
\phi \Big(\Big(\frac{1}{V}
\sum_{x=1}^V a_x^* \Big) \Big( \frac{1}{V} \sum_{y=1}^V a_y
\Big) \Big) = \sup_{N\ge 1}\, \phi \Big( \frac{1}{V^2} P_N^{V}
\sum_{x,y=1}^V a_x^* a_y P_N^{V} \Big), $$ where 
$$ P_N^V =
\bigotimes_{x \in V} P_N^{(x)}. 
$$ 
We now write
\begin{eqnarray*} \lefteqn{ \phi \Big(\Big(\frac{1}{V}
\sum_{x=1}^V a_x^* \Big) \Big( \frac{1}{V} \sum_{y=1}^V a_y
\Big)\Big) = } \\ &=& \sup_{N\ge 1} \frac{1}{V^2} \Big\{
\sum_{x=1}^V \phi \big( P_N^{V} n_x \big) + \sum_{x\neq y} \phi
\big( P_N^{V} a_x^* a_y P_N^{V} \big) \Big\}.
\end{eqnarray*} The first term is clearly bounded by $\frac{1}{V}
\phi(n_x)$ and hence tends to zero. By permutation invariance, the
second term equals $$ \frac{V-1}{V} \phi \left( P_N^{V} a_1^* a_2
P_N^{V} \right). $$ We conclude by proving that the limit (first $ N\to\infty $ and subsequently $ V\to\infty $) of this
expression exists and equals 

\be \phi(a_1^* a_2) = \lim_{N \to
\infty} \phi (P_N^{(1)} a_1^* P_N^{(1)} P_N^{(2)} a_2 P_N^{(2)} ).
\ee 
To this end we write
\begin{eqnarray*} \lefteqn{\phi (P_{N_1}^V a_1^* a_2 P_{N_1}^V)
- \phi (P_{N_2}^V a_1^* a_2 P_{N_2}^V) } \\ &=& \phi \left(
(P_{N_1}^V - P_{N_2}^V) a_1^* a_2 P_{N_1}^V \right) + \phi \left(
P_{N_2}^V a_1^* a_2 (P_{N_1}^V - P_{N_2}^V) \right)
\end{eqnarray*} and treat each term separately. Both terms are
similar; we consider only the first. We have \begin{eqnarray*}
\lefteqn{ |\phi \left( (P_{N_1}^V - P_{N_2}^V) a_1^* a_2 P_{N_1}^V
\right)| } \\ &\leq& \left[ \phi \left( (P_{N_1}^V - P_{N_2}^V)
a_1^* a_1 (P_{N_1}^V - P_{N_2}^V) \right) \right]^{1/2} \left[
\phi \left( P_{N_1}^V a_2^* a_2 P_{N_1}^V \right) \right]^{1/2}.
\end{eqnarray*} The second factor is obviously bounded by
$\phi(n_2)^{1/2}$. In the first factor we can write
\begin{eqnarray*} P_{N_1}^V - P_{N_2}^V &=& (P_{N_1}^{(1)} -
P_{N_2}^{(1)}) \otimes P_{N_1}^{(2)} \otimes \dots \otimes
P_{N_1}^{(V)} \\ && + \dots + P_{N_2}^{(1)} \otimes \dots \otimes
P_{N_2}^{(V-1)} \otimes (P_{N_1}^{(V)} - P_{N_2}^{(V)}).
\end{eqnarray*} With the observation that $P_{N_1}^V - P_{N_2}^V$
commutes with $a_1^* a_1$ we get 
\begin{eqnarray*} \lefteqn{ \phi
((P_{N_1}^V - P_{N_2}^V) a_1^* a_1 (P_{N_1}^V - P_{N_2}^V))} \\
&\leq & \phi ((P_{N_1}^{(1)} - P_{N_2}^{(1)}) a_1^* a_1
(P_{N_1}^{(1)} - P_{N_2}^{(1)})) + (V-1) \phi(P_{N_1}^{(x)} -
P_{N_2}^{(x)}),
\end{eqnarray*} 
and therefore the limit of $ P_N^{V} a_x^* a_y P_N^{V} $ exists.
Similarly, one proves that
$$ \Big|\phi (P_{N_1}^V a_1^* a_2 P_{N_1}^V) - \phi (P_N^{(1)} a_1^*
P_N^{(1)} P_N^{(2)} a_2 P_N^{(2)} )\Big| \to 0 $$  for fixed $V$.
Taking the limit $N \to \infty$ and subsequently $V \to \infty$
the result follows. \qed
\end{proofsect}

\begin{bem} Since $\phi$ is regular, its restriction $\phi_V$
to each ${\mathcal  A}_V$ is regular, and the number operators $n_x$
are well-defined. Moreover, the corresponding GNS representation
is equivalent with the Fock representation by Von Neumann's
theorem \cite{BR}. In particular, $\phi_V$ is normal for all $V$,
i.e. $\phi$ is locally normal. Thus $\phi_V$ has a density matrix
$\rho_{\phi_V}$.
\end{bem}
\medskip

Lemma~\ref{meanenergy} gives the mean energy for our model. Next we are concerned with the relative entropy with respect to our reference state $ \omega_V $. It is well-known that the relative entropy
$$ S(\phi_V\,\|\,\omega_V) = \Tr \left[ \rho_{\phi_V}( \ln
\rho_{\phi_V} - \ln \rho_{\omega_V}) \right] $$ is convex and
superadditive \cite{OP}. A precise definition of
$S(\phi\,\|\,\omega)$ for states on a Von Neumann algebra was
given by Araki \cite{A1, A2}. Given the standard representation
$\pi$ of ${\mathcal  B}({\mathcal  F}_V)$ (the GNS representation with
respect to the tracial state) one has 

\begin{equation} 
S(\phi_V\,\|\,\omega_V)
= -\langle \Phi_V|\,\ln \Delta_{\Omega_V,\Phi_V}\,|\Phi_V \rangle,
\end{equation} where
$$ \Delta_{\Omega_V,\Phi_V} = \pi(\rho_{\omega_V}) \pi'(\rho_{\phi_V})^{-1} $$ 
is the \textit{relative modular operator}. Here $ \Phi_V=\rho_{\phi_V}^{1/2} $ and $ \Omega_V=\rho_{\omega_V}^{1/2} $ respectively, and hence $ \phi_V(A)=\langle \Phi_V|\pi(A)\Phi_V\rangle $ and $ \omega_V(A)=\langle \Omega_V,\pi(A)\Omega_V\rangle $. If $E_{\Omega,\Phi}$ is
the corresponding resolution of the identity, then we can write
\begin{eqnarray} 
S(\phi_V\,\|\,\omega_V) &=& - \int_0^1 \ln
\lambda\, \langle \Phi_V|\,
E_{\Omega_V,\Phi_V}(\d\lambda)\,|\Phi_V \rangle \non \\ && \quad -
\int_1^\infty \ln \lambda\, \langle \Phi_V|\,
E_{\Omega_V,\Phi_V}(\d\lambda)\,|\Phi_V \rangle. \end{eqnarray}
Here we have separated the two integration domains to indicate
that the second integral is always convergent, whereas the first
should be interpreted as $$ -\inf_{\delta > 0} \int_\delta^1 \ln
\lambda\, \langle \Phi_V|\,
E_{\Omega_V,\Phi_V}(\d\lambda)\,|\Phi_V \rangle
$$ and determines whether $S(\phi_V\,\|\,\omega_V)$ is finite or
infinite. Note that $ S(\phi_V\,\|\,\omega_V)=+\infty $ if $ \Phi_V\not\leq \supp\,(\omega_V) $. 

An equivalent definition was introduced by Uhlmann
\cite{Uhlmann}: 

\begin{equation} S(\phi_V\,\|\,\omega_V) = - \lim_{t \downarrow
0} \frac{1}{t} \left(\|\Delta_{\Omega_V,\Phi_V}^{t/2} \Phi_V
\|^2 - 1 \right). 
\end{equation} 
This is easily seen to be equivalent using
Lebesgue's monotone convergence theorem. We now prove the following variational formula for the relative entropy (compare \cite[Lemma~3.2.13]{DS01} for classical probability theory):

\begin{theorem}\label{L2}
For any state $\phi_V$ on ${\mathcal  B}({\mathcal  F}_V)$,
$$ 
\begin{aligned}
S(\phi_V\,\|\, \omega_V) &= \sup_{A \in {\mathcal  B}({\mathcal  F}_V)\colon
A^*=A} \left\{ \beta \phi_V(A) - \ln \Tr {\rm e}^{\beta(\mu {\mathcal  N} -
\sum_{x=1}^V h_x + A)} \right\} \\ & \qquad + V \ln \Tr
{\rm e}^{\beta (\mu n_1-h_1)}. 
\end{aligned}
$$
\end{theorem}

\begin{proofsect}{Proof}
We first prove that $S(\phi_V\,\|\, \omega_V)$ is
greater than the right-hand side. Let $A \in {\mathcal  B}({\mathcal  F}_V)$
be self-adjoint. Then we can define the perturbed state $\psi_V$
by $$ \rho_{\psi_V} = \frac{{\rm e}^{\beta(\mu {\mathcal  N} - \sum_{x=1}^V
h_x + A)}}{\Tr {\rm e}^{\beta(\mu {\mathcal  N} - \sum_{x=1}^V h_x + A)}}. $$
We write $$ Z_{A,V} = \Tr {\rm e}^{\beta(\mu {\mathcal  N} - \sum_{x=1}^V h_x
+ A)} $$ and $ Z_V= \Tr {\rm e}^{\beta(\mu {\mathcal  N} - \sum_{x=1}^V h_x} $ in the following. We employ a change of state (measure) method \cite{DS01} with respect to the reference state $ \omega_V $ and the perturbed reference state $ \psi_V $. The non-commutativity of our random variables is an additional difficulty.
But the Du Hamel formula gives
\begin{equation} 
\begin{aligned}
\langle &\Phi_V|\,\Delta_{\Psi_V,\Phi_V}^t\,|\Phi_V \rangle\\&=
Z_{A,V}^{-t} Z_V^t \Big\langle \Phi_V|\, \Big(\1 - \beta \int_0^t
\d\tau\, \pi ({\rm e}^{-\beta \tau H_V^A} A {\rm e}^{\beta \tau H_V^0}\Big)
\pi(\rho_{\omega_V}^t) \pi'(\phi_V^{-t})|\Phi_V \Big\rangle.
\end{aligned}
\end{equation}
where $$ H_V^A = -\mu {\mathcal  N} + \sum_{x=1}^V h_x - A, $$
and where $ H_V^0=-\mu {\mathcal  N} + \sum_{x=1}^V h_x $ is the non-interacting part of the Hamiltonian in \eqref{Hamiltonian}.
Differentiating we get
\begin{equation} 
\begin{aligned}
\frac{\d}{\d t}\big|_{t=0^+} \langle
\Phi_V|\,\Delta_{\Psi_V,\Phi_V}^t\,|\Phi_V \rangle &=
-\ln Z_{A,V} +\ln Z_V- \beta\langle \Phi_V|\,\pi(A) \Phi_V\rangle\\ 
& + \frac{\d}{\d t}\big|_{t=0^+} \langle
\Phi_V|\,\Delta_{\Omega_V,\Phi_V}^t\,|\Phi_V \rangle
\end{aligned}
\end{equation} and hence
\begin{equation}\label{rentid}
S(\phi_V\,\|\, \psi_V) = S(\phi_V\,||\, \omega_V) - \beta
\phi_V(A) + \ln \frac{\Tr {\rm e}^{\beta(\mu {\mathcal  N} - \sum_{x=1}^V h_x
+ A)}}{\Tr {\rm e}^{\beta(\mu {\mathcal  N} - \sum_{x=1}^V h_x)}}.
\end{equation} 

The desired inequality now follows from the
positivity of the relative entropy. (Notice that this follows
immediately from $\ln \lambda \leq 1-\lambda$ and
$$
\begin{aligned}
 \int \lambda \langle
\Phi_V|\,E_{\Psi_V,\Phi_V}(\d\lambda)\,|\Phi_V \rangle &= \langle
\Phi_V|\,\Delta_{\Psi_V,\Phi_V}\,|\Phi_V \rangle \non \\ &=
\langle \Phi_V|\,\pi(\rho_{\psi_V})
\pi'(\rho_{\phi_V})^{-1}\,|\Phi_V \rangle = \Tr (\rho_{\psi_V}) =
1 
\end{aligned} 
$$
by a simple approximation.)
\medskip

To prove the converse inequality, first assume that $c_1 \omega_V
\leq \phi_V \leq c_2 \omega_V$ for constants $0 < c_1 < c_2 <
+\infty$. Then there exists a bounded relative Hamiltonian $A$
such that 
$$ \rho_{\phi_V} = \frac{{\rm e}^{\beta(\mu {\mathcal  N} -
\sum_{x=1}^V h_x + A)}}{\Tr {\rm e}^{\beta(\mu {\mathcal  N} - \sum_{x=1}^V
h_x + A)}}, $$ 
that is,
$$ \rho_{\phi_V} = \frac{{\rm e}^{-\beta H_V^A}}{\Tr {\rm e}^{-\beta H_V^A}}
\;\mbox{ and } \;\rho_{\omega_V} = \frac{{\rm e}^{-\beta (H_V^A + A)}}{\Tr
{\rm e}^{-\beta (H_V^A + A)}}. 
$$ 

\noindent Indeed, it follows easily that ${\rm
Dom}(\ln \rho_{\phi_V}) = {\rm Dom}(\ln \rho_{\omega_V})$ and $A =
\ln \rho_{\phi_V} - \ln \rho_{\omega_V} $ is bounded. The identity
(\ref{rentid}) with $\phi_V = \psi_V$ then yields 
\begin{equation}\label{entropyid}
S(\phi_V\,\|\, \omega_V) = \beta \phi_V(A) - \ln \Tr {\rm e}^{\beta(\mu
{\mathcal  N} - \sum_{x=1}^V h_x + A)} + V \ln \Tr {\rm e}^{\beta (\mu
n_1-h_1)}.
\end{equation}

The general case then follows from the lower semi continuity of the
relative entropy \cite{OP}:
$$ S(\phi_V\,\|\, \omega_V) \leq \liminf_{V \to \infty}
S(\phi_{V,\eps}\,\|\, \omega_V). $$ Indeed, first assuming $\phi_V
\leq \lambda \omega_V$ we can put $\phi_{V,\eps} = (1-\eps) \phi_V
+ \eps \omega_V$ to conclude that the theorem holds in this case.
In the general case, we use the approximation
$\rho_{\phi_{V,\eps}} = \frac{P_\eps \rho_{\phi_V}
P_\eps}{\Tr[P_\eps \rho_{\phi_V} P_\eps]}. $ 
\qed
\end{proofsect}

It is proved in \cite{OP}, Corollary 5.21, that \begin{equation}
S(\phi\,\|\,\omega_1 \otimes \omega_2) \geq
S(\phi_1\,\|\,\omega_1) + S(\phi_2\,\|\,\omega_2).
\end{equation} The relative entropy is therefore superadditive:
\begin{equation} S(\phi_{V1+V2}\,\|\,\omega_{V_1} \otimes \omega_{V_2}) \geq
S(\phi_{V_1}\,\|\,\omega_{V_1}) + S(\phi_{V_2}\,\|\,\omega_{V_2}).
\end{equation} It follows that the mean entropy
$$ 
s(\phi\,\|\,\omega):= \lim_{V \to \infty} \frac{1}{V}
S(\phi_V\,\|\,\omega_V) 
$$ exists. We now have the following \lq
level-III' variational expression for the pressure. Here level-III refers to the fact that in the variational formula we optimise over states of the infinite system \cite{georgii}.
The set of all regular translation-invariant states on $ \mathcal A $ is denoted by $ \Scal (\Acal) $, and the set of all regular translation-invariant and permutation-invariant states on $ \Acal $ by $ \Scal_\Pi(\Acal) $.

\begin{theorem}\label{Th1}

$$ 
\begin{aligned}p(\beta, \mu, \lambda) &:=
\lim_{V \to \infty} \frac{1}{\beta V} \ln \Tr {\rm e}^{\beta(\mu {\mathcal 
N} - H_V)} \\ &= \sup_{\heap{\phi\in\Scal_\Pi\colon}{\phi(n_x) < +\infty}} \Big\{
\phi(a_1^* a_2) - \frac{1}{\beta} s(\phi\,\|\, \omega) \Big\}
+ \frac{1}{\beta} \ln \Tr {\rm e}^{\beta(\mu n_1 - h_1)}.
\end{aligned} 
$$

\noindent Here the supremum is taken over all regular translation- and permutation-invariant states $\phi$ on $\mathcal  A$ such that
$\phi(n_x) < +\infty$ for all $x \in \N$. 
\end{theorem}

\begin{proofsect}{Proof}
We denote 
$$ v_V = \Big( \frac{1}{V} \sum_{x=1}^V
a_x^* \Big) \Big( \frac{1}{V} \sum_{y=1}^V a_y\Big) 
$$ and
$$ P_V = \frac{1}{\beta} \ln \Tr {\rm e}^{\beta (\mu {\mathcal  N} -
\sum_{x=1}^V h_x + V v_V)} $$ so that $$ p(\beta, \mu, \lambda) =
\lim_{V \to \infty} \frac{1}{V} P_V. $$ 
We approximate $v_V$ using a
cut-off, and call this bounded operator also $ v_V $.
By Theorem~\ref{L2}
$$ 
S(\phi_V\,\|\, \omega_V) \geq \beta V \phi_V(v_V) - \ln \Tr
{\rm e}^{\beta(\mu {\mathcal  N} - \sum_{x=1}^V h_x + V v_V)}
+ V \ln \Tr {\rm e}^{\beta (\mu n_1-h_1)}. 
$$  
This implies,  using Lemma~\ref{L1}, that
$$
\begin{aligned}
\liminf_{V \to \infty} \frac{1}{V} P_V &\geq  \sup_{\phi:\, \phi(n_x) < +\infty} \Big\{\phi(a_1^* a_2) -
\frac{1}{\beta} s(\phi\,\|\, \omega)\Big\}\\
& + \frac{1}{\beta} \ln \Tr {\rm e}^{\beta(\mu n_1 - h_1)}.
\end{aligned}
$$

To prove the converse we seek an approximate maximiser. This is
standard. We let $${\tilde \psi}_V = \psi_V \otimes \psi_V \otimes
\dots $$ be the infinite tensor product of states and define $$
{\bar \psi}_V = \frac{1}{V} \sum_{j=1}^V {\tilde \psi}_V \circ
\tau_{j-1}, $$ where $\tau_j$ is the translation over $j$. This is
a permutation-invariant state. We estimate the expectation of the
energy density: 
\begin{equation} 
\begin{aligned}
{\bar \psi}_V (a_1^* a_2)&=
\frac{1}{V} \sum_{k=1}^V {\tilde \psi}_V(a_k^* a_{k+1}) \\
&= \frac{1}{V} \sum_{k=1}^{V-1} \psi_V(a_k^* a_{k+1}) +
\frac{1}{V} \psi_V(a_V^*) \psi_V(a_1) \\ &= \frac{1}{V^2}
\sum_{x=1}^V \sum_{y=1;\, y\neq x}^V \psi_V(a_x^* a_y) +
\frac{1}{V} \psi_V(a_V^*) \psi_V(a_1) \\ &=\frac{1}{V^2}
\sum_{x,y=1}^V \psi_V(a_x^* a_y) - \frac{1}{V} \psi_V(a_1^* a_1) +
\frac{1}{V} | \psi_V(a_1)|^2. 
\end{aligned}
\end{equation} 
Lemma~\ref{L1} and the
Cauchy-Schwarz inequality $|\psi_V(a_1)|^2 \leq \psi_V(a_1^* a_1)$
then imply that 
\begin{equation} |{\bar \psi}_V(a_1^* a_2) -
\psi_V(v_V)| \to 0 \mbox{ as } V\to\infty. 
\end{equation}

It is known that the entropy is convex in both arguments
\cite{OP}. In particular, we have
$$ 
s(\lambda \phi_1 + (1-\lambda) \phi_2\,\|\, \omega) \leq
\lambda s(\phi_1\,\|\, \omega) + (1-\lambda) s(\phi_2 \,\|\,
\omega). 
$$ 
On the other hand, by a simple approximation, we have
$$
\begin{aligned}
 S(\lambda \phi_{1,V}+ (1-\lambda) \phi_{2,V}\,\|\, \omega_V) =&-
\lambda \langle \Phi_{1,V} |\, \ln \Delta_{\Phi_V,\Omega_V} \,|
\Phi_{1,V} \rangle\\ & - (1-\lambda) \langle \Phi_{2,V} |\,
\ln \Delta_{\Phi_V,\Omega_V} \,| \Phi_{2,V} \rangle,
\end{aligned} 
$$
where $\phi_V = \lambda \phi_{1,V} + (1-\lambda)
\phi_{2,V}$. Using the fact that $\pi(\rho_{\Omega_V})$ and
$\pi'(\rho_{\phi_V})$ respectively $\pi'(\rho_{\phi_{1,V}}),\pi'(\rho_{\phi_{2,V}}) $, commute
and the operator monotonicity of the inverse and the
logarithm, we have 
\begin{eqnarray*} \pi'(\rho_{\phi_V}) \geq
\lambda \pi'(\rho_{\phi_{1,V}}) &\implies &
\Delta_{\Omega_V,\Phi_V} \leq \lambda^{-1}
\Delta_{\Omega_V,\Phi_{1,V}} \non \\ &\implies & \ln
\Delta_{\Omega_V,\Phi_V} \leq \ln \Delta_{\Omega_V,\Phi_{1,V}} -
\ln \lambda. \end{eqnarray*} Of course, a similar inequality holds
w.r.t. $\phi_{2,V}$. Therefore
$$
\begin{aligned}
S(\lambda \phi_{1,V}+ (1-\lambda) \phi_{2,V}\,\|\, \omega_V)
 & \geq  \lambda S(\phi_{1,V}\,\|\, \omega_V) + (1-\lambda)
S( \phi_{2,V}\,\|\, \omega_V) \\
& \quad + \lambda \ln \lambda +  (1-\lambda) \ln (1-\lambda).
\end{aligned}
$$
In the limit, we find in combination with the convexity above,
that the mean relative entropy is affine in the first variable:
\begin{equation} s(\lambda \phi_{1}+ (1-\lambda) \phi_{2}\,\|\, \omega)
= \lambda s(\phi_{1}\,\|\, \omega_V) + (1-\lambda) s(
\phi_{2}\,\|\, \omega). \end{equation} Applying this to the state
${\bar \psi}_V$ we get $$ s({\bar \psi}_V\,\|\,\omega) =
\frac{1}{V} \sum_{k=1}^V s({\tilde \psi}_V \circ \tau_{k-1}
|\, \omega) $$ provided the right-hand side exists. However, by
the translation-invariance of $\omega$, the right-hand side can be
written as
$$
\begin{aligned}
\frac{1}{V} \sum_{k=1}^V
s({\tilde \psi}_V \circ \tau_{k-1} \| \, \omega) &= \lim_{n
\to \infty} \frac{1}{nV^2} \sum_{k=1}^V S((\psi_V)^{\otimes n}
\circ \tau_{k-1} \|\, \omega_{nV}) \\ &= \lim_{n \to \infty}
\frac{1}{nV} S((\psi_V)^{\otimes n}\,\|\, (\omega_V)^{\otimes n})
\\ &= \frac{1}{V} S(\psi_V\,\|\, \omega_V). 
\end{aligned} 
$$
We
therefore have
\begin{equation}
\Big({\bar\psi}_V (a_1^* a_2) - \frac{1}{\beta} s({\bar\psi}_V\,\|\, \omega)\Big) - \Big(\psi_V(v_V) -
\frac{1}{\beta V} S(\psi_V\,\|\,\omega_V)\Big) \to 0
\end{equation}  as $V \to \infty$. On the other hand, by
Theorem~\ref{L2} as above (see Eq. (\ref{entropyid})),
\begin{equation} 
\begin{aligned}
S(\psi_V\,\|\, \omega_V) &=\beta V \psi_V(v_V)
- \ln \frac{\Tr {\rm e}^{\beta(\mu {\mathcal  N} - \sum_{x=1}^V h_x + V
v_V)}}{\Tr {\rm e}^{\beta V (\mu n_1-h_1)}}\\ 
&=\beta\psi_V(v_V)- \beta P_V 
\beta V + \ln \Tr {\rm e}^{\beta (\mu n_1-h_1)}.
\end{aligned}
\end{equation} 
\qed
\end{proofsect}

The variational expression for the pressure can now be simplified
by decomposing the state $\phi$ into an integral of extremal
permutation-invariant states:
\begin{theorem} 
\begin{equation} 
\begin{aligned}
p(\beta, \mu, \lambda)
&=\sup_{\heap{\sigma \in {\mathcal  S}({\mathcal  A}_1)\colon}{\sigma(n_1) <
+\infty}} \Big\{|\sigma(a_1)|^2 - \frac{1}{\beta} S(\sigma\,\|\,
\omega_1) \Big\}
+ \frac{1}{\beta} \ln \Tr {\rm e}^{\beta(\mu n_1 - h_1)},
\end{aligned}
\end{equation} where the supremum is now taken over regular states $\sigma$ of
${\mathcal  A}_1$ such that $\sigma(n_1) < +\infty$. \label{Th2}
\end{theorem}

\begin{proofsect}{Proof}
The set of permutation invariant states ${\mathcal S}_\pi({\mathcal  A})$ is a convex compact set in the weak$*$-topology,
and it is metrizable because ${\mathcal  A}$ is separable. By Choquet's
theorem \cite{Phelps}, we can therefore decompose an arbitrary
state $\phi$ into an integral 
$$ \phi = \int_{{\rm ext}\,({\mathcal 
S}_\pi({\mathcal  A}))} \psi \,\mu(\d\psi) $$ over the extremal points of
${\mathcal  S}_\pi({\mathcal  A})$. Here $\mu$ is a probability measure on
${\rm ext}\,({\mathcal  S}_{\pi}({\mathcal  A}))$. But, by St{\o}rmer's theorem
\cite{Stormer}, the extremal permutation invariant states are the
product states
$$ \psi_\sigma = \sigma \otimes \sigma \otimes \dots, $$ where
$\sigma$ is a state of ${\mathcal  A}_1$. Thus, 
$$ \phi = \int_{{\mathcal 
S}({\mathcal  A}_1)} \psi_\sigma\, \mu(\d\sigma). 
$$ Moreover, since
$\phi(n_x) < \infty$, we have that $\sigma(n_1) < \infty$ for
$\mu$-almost every $\sigma$. It follows that 
$$ \phi(a_1^* a_2) =\int \sigma(a_1^*) \sigma(a_1)\, \mu(\d\sigma) = \int |\sigma(a_1)|^2\,
\mu(\d\sigma). $$ For the entropy term we use the following lemma
\cite[Lemma~9.7]{Phelps}:

\begin{lemma} Suppose that $X$ is a compact convex subset of a
locally convex topological vector space. Let $f\colon X \to \R$ be an
affine, lower semi continuous function on $X$, and suppose that
$\mu$ is a (Radon) probability measure on $X$, $x_0 = \int x\,
\mu(dx)$. Then $$ \int f(x)\, \mu(\d x) = f(x_0). $$ \label{L3}
\end{lemma}\medskip

Since the relative entropy (and hence the mean relative entropy)
is lower semi continuous and affine, the lemma applies and we have
$$ 
s(\phi\,\|\, \omega) = \int s(\psi_\sigma\,\|\,\omega)
\,\mu(\d\sigma). 
$$ 
However, since $\omega$ is also a product
measure, $s(\psi_\sigma\,\|\,\omega) = S(\sigma\,\|\omega_1).$
The Theorem now follows. 
\qed
\end{proofsect}

\begin{bem} Notice that the subset of regular states is also
closed in the set of all states on ${\mathcal  A}$ by the
Banach-Steinhaus theorem. Indeed, if $\phi_\alpha$ is a net of
regular states converging to $\phi$ in weak$*$-topology then
$\phi_\alpha(W(tf))$  converges uniformly on compact sets $t \in
[a,b]$, where $W(tf)$ is the Weyl operator for $f \in \C^V$.
\end{bem}
\medskip

In the following we write $a$ instead of $a_1$ and $n$ for $n_1$.
Our variational expression for the pressure  can be further reduced to

\begin{theorem}\label{Th3} 
\begin{equation}
p(\beta,\mu,\lambda) = \sup_{z \in \C} \{|z|^2 - I(z)\} +
\frac{1}{\beta} \ln \Tr {\rm e}^{\beta(\mu n - h)}, \end{equation} where
the rate function $I(z)$ is given by
\begin{equation} I(z) = \sup_{\nu \in \C} \Big\{ {\bar \nu}z +
\nu {\bar z} - \frac{1}{\beta} \ln \frac{\Tr {\rm e}^{\beta (\mu n - h +
(\nu a^* + {\bar \nu} a))}}{\Tr {\rm e}^{\beta (\mu n - h)}} \Big\}.
\end{equation} 
\end{theorem}

To prove this, we first need a lemma:

\begin{lemma}\label{L4} Denote 
\begin{equation}
{\tilde p}(\nu) = \frac{1}{\beta} \ln \Tr {\rm e}^{\beta (\mu n - h +
(\nu a^* + {\bar \nu} a))}. 
\end{equation} This function is convex
and satisfies 
\begin{equation} {\tilde p}(\nu) \sim |\nu|^{4/3} \mbox{ as }
|\nu| \to \infty. 
\end{equation}
\end{lemma}

\begin{proofsect}{Proof of Lemma~\ref{L4}} Differentiating, we have $$ {\tilde p}'(\nu) =
\langle a^* \rangle_{H(\nu)}, $$ where $$ H(\nu) = -\mu n + h -
\nu a^* - {\bar \nu} a. $$ Differentiating again, we get, as in
\cite{BD}, with $\nu = x+iy$, 
$$ 
\frac{\partial^2}{\partial x^2}{\tilde p}(\nu) = \Big(\frac{\partial}{\partial \nu} +
\frac{\partial}{\partial {\bar \nu}}\Big)^2 {\tilde p}(\nu) =
\beta\big(a+a^* -\langle a+a^* \rangle\,|\,a+a^* - \langle
a+a^* \rangle \big)_{H(\nu)}, $$ $$ \frac{\partial^2}{\partial
y^2} {\tilde p}(\nu) = -\Big( \frac{\partial}{\partial \nu} -
\frac{\partial}{\partial {\bar \nu}} \Big)^2 {\tilde p}(\nu) =
\beta \big( a-a^* - \langle a-a^* \rangle\,|\,a-a^* - \langle
a-a^* \rangle \big)_{H(\nu)}, 
$$ 
and 
$$
\begin{aligned}
\frac{\partial^2}{\partial x \partial y} {\tilde p}(\nu) &=
i\Big(\frac{\partial}{\partial \nu} + \frac{\partial}{\partial
{\bar \nu}}\Big) \Big(\frac{\partial}{\partial \nu} -
\frac{\partial}{\partial {\bar \nu}}\Big){\tilde p}(\nu) 
\\ &= i\beta \big( a+a^* - \langle a+a^* \rangle\,|\,a^*-a - \langle
a^*-a \rangle \big)_{H(\nu)}. 
\end{aligned} 
$$
It follows by the
Cauchy-Schwarz inequality that the corresponding matrix is
positive-definite.

\noindent To prove the asymptotic behaviour, we first remark that $$ \nu a^*
+ {\bar \nu} a \leq 2 |\nu| (n+1)^{1/2} $$ and hence

$$
\begin{aligned}{\tilde p}(\nu) &\leq\frac{1}{\beta} \ln \Tr
{\rm e}^{\beta(\mu n - h + 2 |\nu| (n+1)^{1/2})} \\ &=
\frac{1}{\beta} \ln \sum_{n=0}^\infty {\rm e}^{\beta((\mu-1)n - \lambda
n (n-1) + 2 |\nu| \sqrt{n+1})} = O(|\nu|^{4/3}). 
\end{aligned}
$$
For the reverse inequality we use one half of the Berezin-Lieb
bounds \cite{Lieb}:
$$ \Tr {\rm e}^{-H(\nu)} \geq \int \frac{\d z\,\d {\bar z}}{\pi} {\rm e}^{-\langle
z \,|\,H(\nu)\,|\,z\rangle}. $$ Here $|z\rangle$ stands for the
coherent state $$ |z\rangle = {\rm e}^{-\half |z|^2} \sum_{n=0}^\infty
\frac{z^n (a^*)^n}{n!} |0\rangle. $$ Since $$ \langle
z|\,a\,|z\rangle = z, \hskip 20pt \langle z|\,a^*\,|z\rangle =
{\bar z}, $$
$$ \langle z|\,n\,|z\rangle = |z|^2, $$ and $$ \langle
z|\,(a^*)^2 a^2\,|z\rangle = |z|^4, $$ we get $$ \Tr {\rm e}^{-H(\nu)}
\geq \int \frac{\d z\,\d {\bar z}}{\pi}\, {\rm e}^{\beta ((\mu-1)|z|^2 -
\lambda |z|^4 + {\bar \nu}z + \nu {\bar z})} = O(|\nu|^{4/3}). $$
\qed
\end{proofsect}

\begin{proofsect}{Proof of Theorem~\ref{Th3}} Take $A$ to be an approximation of
$\nu a^* + {\bar \nu} a$ in Theorem~\ref{L2}. We get, writing 
$$
p(\nu) = \frac{1}{\beta} \ln \Tr {\rm e}^{\beta (\mu n - h + (\nu a^* +
{\bar \nu} a))}, $$ 
the inequality
$$ 
|\phi(a)|^2 - \frac{1}{\beta}
S(\phi_V\,\|\,\omega_V) \leq |\phi(a)|^2 - \phi(\nu a^* + {\bar \nu}
a) + p(\nu) - p(0) $$ and since $\nu$ is arbitrary, $$ |\phi(a)|^2
- \frac{1}{\beta} S(\phi\,\|\,\omega) \leq |\phi(a)|^2 -
I(\phi(a)). $$

Conversely, suppose that $z$ is a maximiser for $\sup \{ |z|^2 -
I(z)\}$. For any $\nu \in \C$, define the state $\phi_\nu$ by
$$ \rho_{\phi_\nu} = \frac{{\rm e}^{\beta (\mu n - h + (\nu a^* +
{\bar \nu} a))}}{\Tr {\rm e}^{\beta (\mu n - h + (\nu a^* + {\bar \nu}
a))}} $$ and choose $\nu$ such that $\phi_\nu(a) = z$. It follows
from the above lemma that such $\nu$ exists. Then
$$
\begin{aligned} 
\sup_{\phi} \Big\{ |\phi(a)|^2 -
\frac{1}{\beta} S(\phi\,\|\,\omega) \Big\} &\geq
|\phi_\nu(a)|^2 - S(\phi_\nu\,||\, \omega) \\ 
&= |z|^2 - \phi_\nu(\nu a^* + {\bar \nu} a) + p(\nu) - p(0) \\ 
&= |z|^2 - I(z).
\end{aligned}
$$ 
\qed
\end{proofsect}

We finally rewrite the expression in the form \eqref{pformula}. With a gauge
transformation it is easy to see that $I(z)$ only depends on $|z|$
and we have 
\begin{equation}\label{var} 
p(\beta, \mu,\lambda) = \sup_{x \geq 0} \left\{
x^2 - I(x) \right\} + p(0)  
\end{equation}
and 
\begin{equation}\label{rate} 
I(x) = \sup_{r\geq 0} \{2rx - p(r)\} + p(0). 
\end{equation}

\noindent Now let $r \geq 0$ be given, and suppose that $x_{\rm max}$ is a
maximiser of the expression (\ref{var}). Then
$$ p(\beta,\mu,\lambda) = x_{\rm max}^2 - I(x_{\rm max}) + p(0)
\leq x_{\rm max}^2 - 2rx_{\rm max} + p(r) $$ and choosing
$r=x_{\rm max}$, $$ p(\beta,\mu,\lambda)  \leq -x_{\rm max}^2  +
p(x_{\rm max})  \leq \sup_{r \geq 0} \{ -r^2 + p(r)\}.$$

\noindent On the other hand, if $r_0$ is a maximiser of the right-hand side,
then
$$ r_0 = \frac{1}{2} \frac{\d}{\d r} p(r) \big\vert_{r=r_0} $$ and
hence
$$ I(r_0) = 2r_0^2 - p(r_0) + p(0). $$ Inserting, we get
$$ \sup_{r \geq 0} \{-r^2 + p(r)\} = -r_0^2 + p(r_0) = r_0^2 -
I(r_0) + p(0) \leq p(\beta,\mu,\lambda). $$
Hence, we have shown the formula \eqref{pformula} for the infinite range Bose-Hubbard model \eqref{Hamiltonian}.

\section{Analysis of the phase diagram}\label{sec-3}

The phase diagram of the model was analysed in \cite{BD}. The same
model, but with disorder, was analysed in \cite{DPZ} where it was
found that the disorder gives rise to new phenomena.
Unfortunately, \cite{BD} contains a few errors, which we wish to
correct here. First of all, the critical values of lambda are not
given by (2.14) of \cite{BD}, but instead \be \lambda_{c,k} =
2k+1. \ee This was already remarked in \cite{DPZ}, see Remark 4.1.
Indeed, although a gap exists for $\lambda > \lambda_k$ given by
(2.14) in \cite{BD}, the limiting value of $\mu(\beta,\lambda)$
lies in this gap only if $\lambda > \lambda_{c,k}$.

There was also a mistake in the program to compute the $p-V$
diagrams of Fig. 5 and 6 in \cite{BD}, as well as the condensation
fractions of Fig. 7 and 8. We include corrected graphs below:
\newpage
\begin{center}
\includegraphics[width=12cm, height=8.5cm]{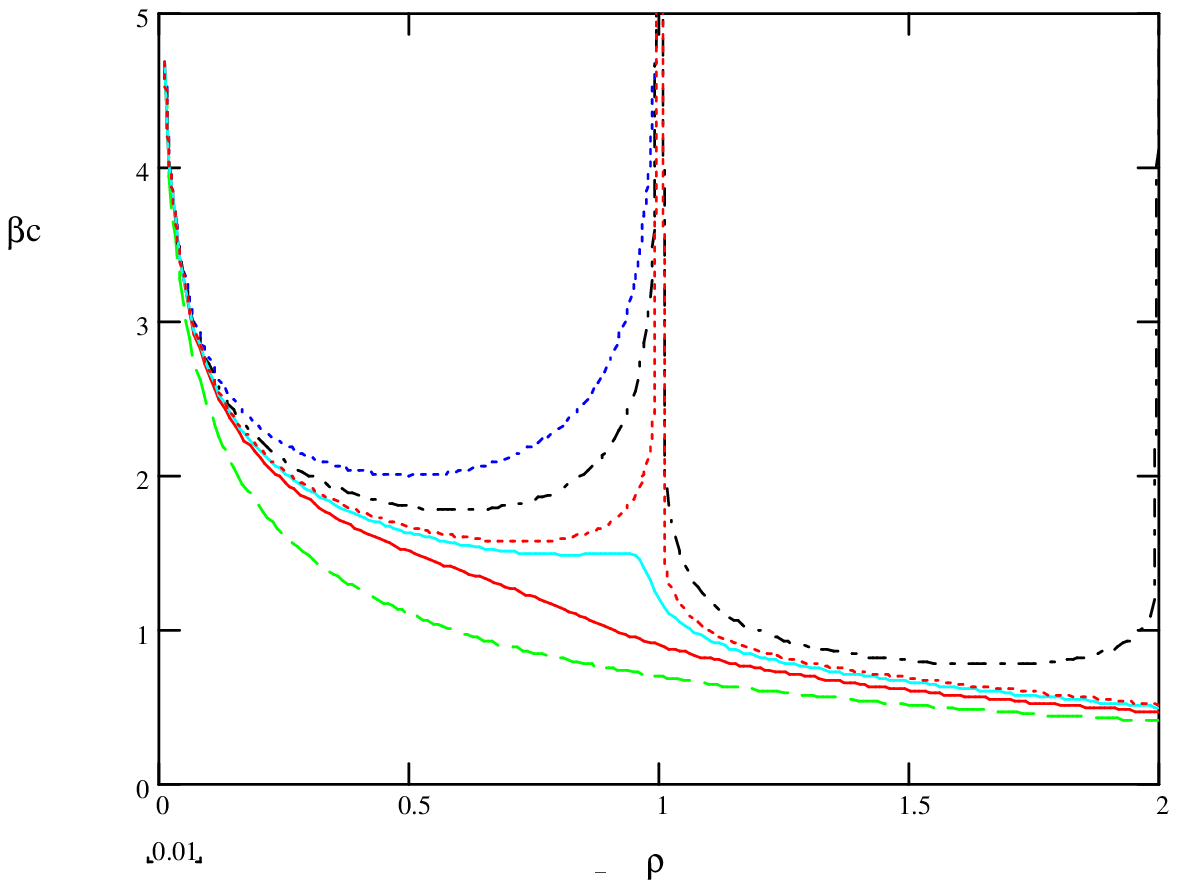}
\label{fig1}
\end{center}
\vskip1cm

\centerline{\emph{The critical inverse temperature as a function}}
\centerline{\emph{of the density for a number of values of
$\lambda$.}}

\begin{center}
\resizebox{12truecm}{7.5truecm}{\includegraphics{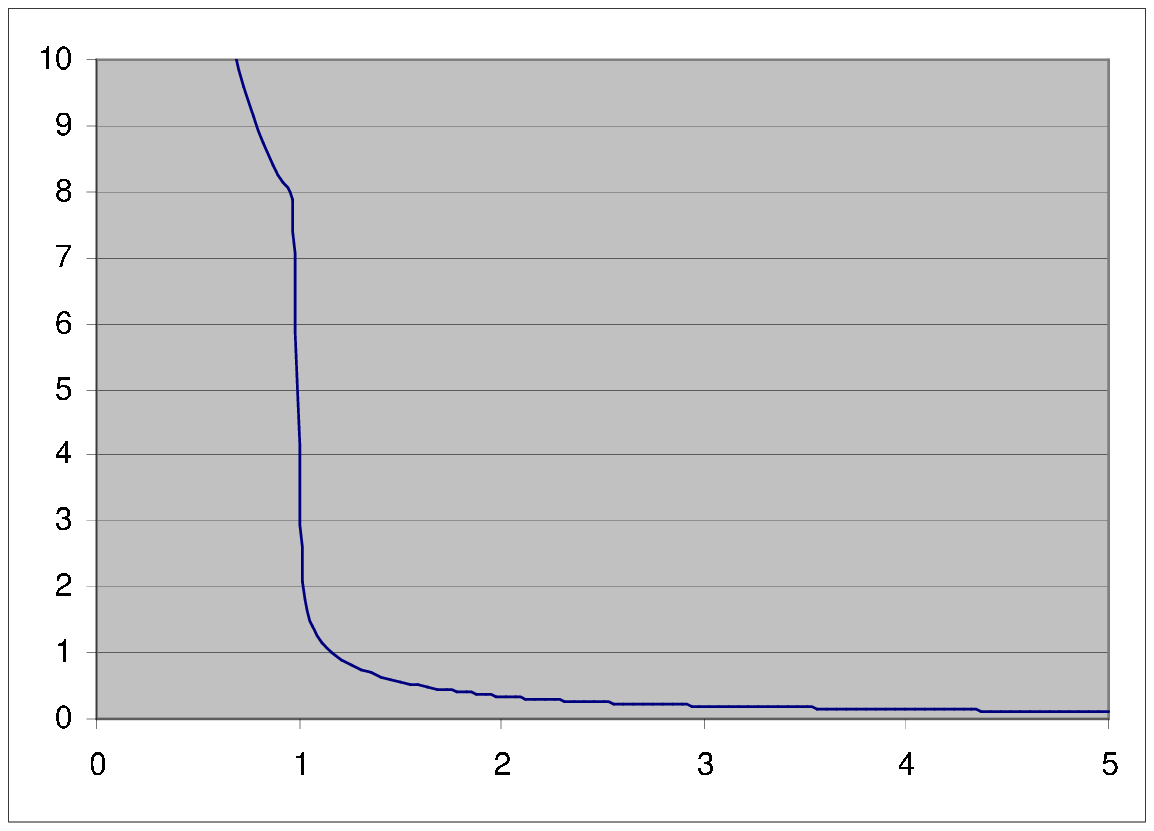}}
\label{fig2}
\end{center}
\vskip1cm

\centerline{\emph{The P-V diagram for $\lambda = 5$.} }

\begin{center}
\resizebox{12truecm}{8truecm}{\includegraphics{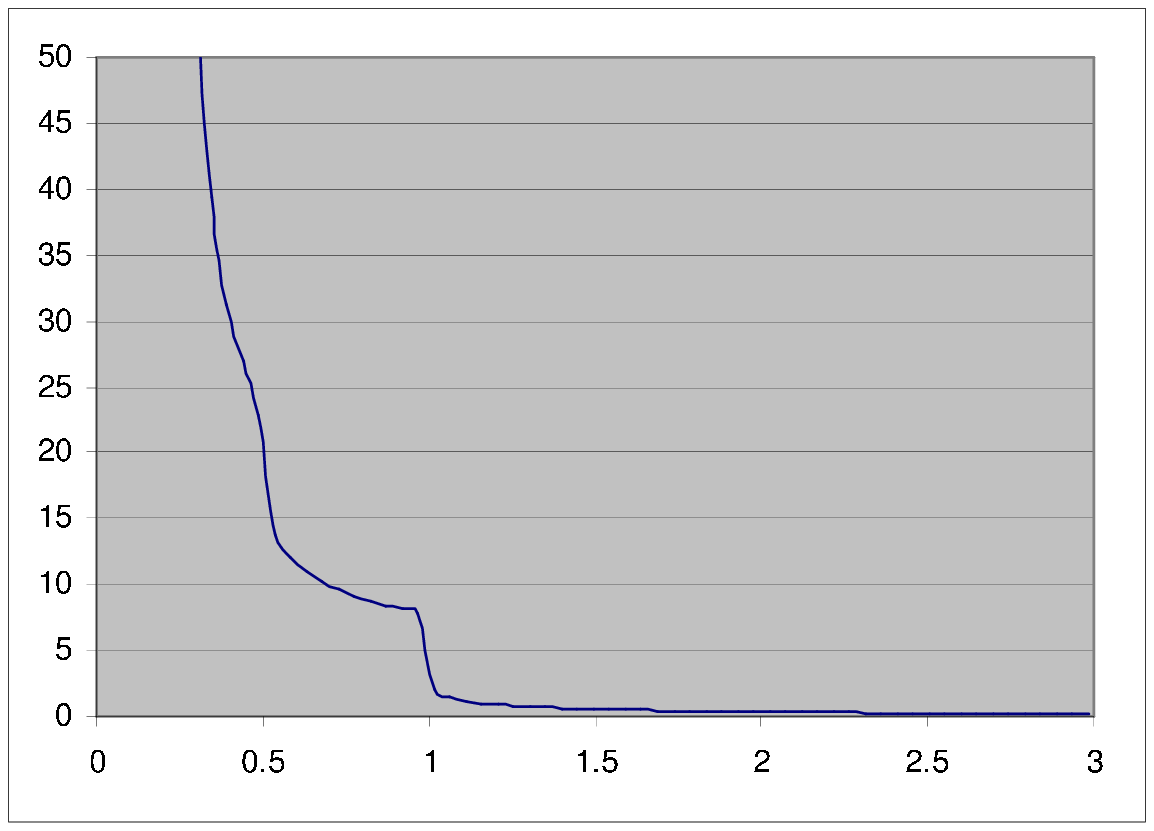}}
\label{fig3}
\end{center}
\vskip1cm

\centerline{\emph{The P-V diagram for $\lambda = 5$ at higher
pressures.}}

\begin{center}
\includegraphics[width=12cm,height=8cm]{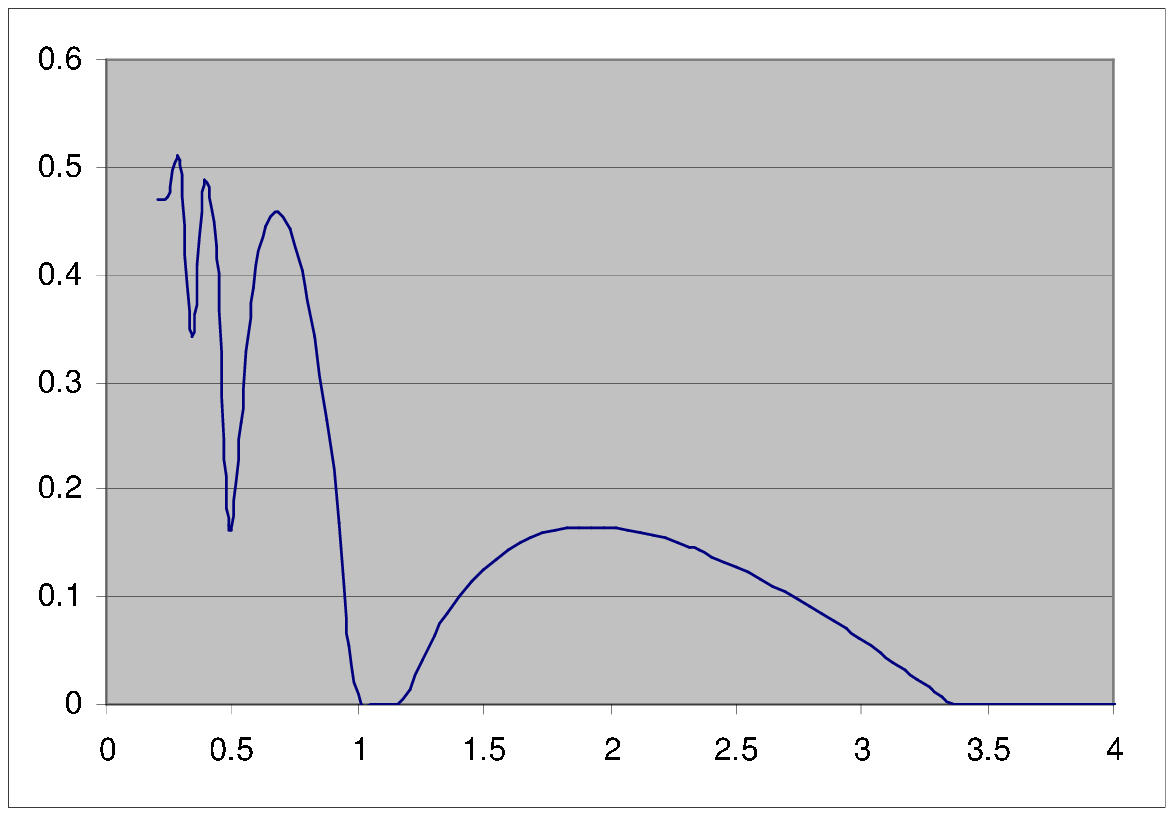}
\label{fig4}
\end{center}

\centerline{\emph{The condensation fraction as a function of the
density for $\lambda = 5$.} }

\newpage
\section{The nearest-neighbour hopping model}\label{sec-4}

As in the case of quantum spin models, there is also a variational
formula for the pressure of the translation-invariant Bose-Hubbard
model, analogous to Theorem~\ref{Th1}:

\begin{theorem} The pressure of the nearest-neighbour hopping Bose-Hubbard
model is given by
\begin{equation} 
\begin{aligned}
p(\beta, \mu, \lambda) &=\sup_{\phi:\,
\phi(n_x) < +\infty} \Big\{ \sum_{\nu = 1}^d \phi(a_0^* a_{{\rm e}_\nu}
+ a^*_{{\rm e}_\nu} a_0) - \frac{1}{\beta} s(\phi\,\|\, \omega)
\Big\}
\\ & \qquad + \frac{1}{\beta} \ln \Tr {\rm e}^{\beta(\mu {\hat n} - h_0)},
\end{aligned}
\end{equation}
where the supremum is over all regular translation-invariant states $\phi$
on the CCR algebra such that $\phi(n_x) < +\infty$, and $\omega$
is the product state $\omega = \bigotimes_{x \in \Z^d} \omega_x$
with
$$ \rho_{\omega_x} = \frac{1}{Z_0} {\rm e}^{\beta(\mu {\hat n}_x - h_x)} $$ and
$$ h_x = d\, {\hat n}_x - \lambda {\hat n}_x({\hat n}_x-1). $$
\end{theorem}

The derivation of this formula is completely analogous to that of
Theorem~\ref{Th1}. The main difference is that the infinite-volume
limit now has to taken in the sense of Van Hove. For the case of
spin models, see for example \cite{Israel} or \cite{Hugenholtz}.
This variational formula does not seem to have been written down
before, though it has to be said that it is not clear how useful
this formula is. The analogous formula for spin models has so far
not been very useful for analysing the phase diagram. One possible
application is perhaps the cluster variation approximation, see
\cite{Kik1}, \cite{Kik2}, \cite{Surda},
\cite{Schlijper}.


\end{document}